\documentclass[aps,pre]{revtex4} 
\usepackage[apple mac]{inputenc} 
\usepackage{graphicx} 
\usepackage[french]{babel} 
\usepackage{picins} 
\usepackage{epsf} 
\usepackage{subfigure} 
\usepackage{amsmath} 
\usepackage{amssymb} 
\usepackage{amsfonts}
\usepackage{bm}
\usepackage{natbib}
\usepackage{epstopdf}
\begin{document}

\title{\bf Quantum graphs and the integer quantum Hall effect} 

\author{ N. Goldman and P. Gaspard} 
\affiliation{Center for Nonlinear Phenomena and Complex Systems,\\ 
Universit\'e Libre de Bruxelles, Code Postal 231, Campus Plaine, B-1050 Brussels, Belgium} 

\begin{abstract} 
We study the spectral properties of infinite rectangular quantum graphs in the presence of a magnetic field. We study how these properties are affected when three-dimensionality is considered, in particular, the chaological properties. We then establish the quantization of the 
Hall transverse conductivity for these systems.
This quantization is obtained by relating the transverse conductivity
to topological invariants. The different integer values of the Hall conductivity are explicitly computed
for an anisotropic diffusion system which leads to fractal phase diagrams. 
\end{abstract} 

\maketitle 

\section{Introduction} 
Quantum graphs have been the focus of much interest  during the last thirty years \cite{alex,degennes,zur}. These models which describe the propagation of a quantum wave within an arbitrary complex object are extremely versatile  allowing the study of various interesting quantum phenomena. Quantum graphs appear in various fields such as solid state physics, quantum chemistry, chaology and wave physics. Basically, quantum graphs describe wave propagation through fine structures. In the field of quantum chemistry, graphs have been used to represent $\pi$-electronic orbitals in organic molecules formed with double chemical bonds  \cite{pauling}. In nanotechnology as for future quantum computer devices, they modelize thin conductor circuits that propagate information. They also describe fine superconducting circuits and wave guides leading to acoustic, optic and electromagnetic applications. Generally speaking, graphs constitute useful models for the description of quantum transport on connected systems \cite{zur}.

In the context of quantum chaology, graphs have been the vehicle to confirm important conjectures about chaos signatures.  Kottos and Smilansky discovered that periodic-orbit theory exactly applies to quantum graphs and that their spectra may obey the Wigner level spacing statistics under certain conditions \cite{kottos}. A minimum of three incommensurate bond lengths is required for Wigner level repulsion to already manifest itself \cite{BG.JSP}. Furthermore, graphs are simple models of quantum scattering and a semiclassical bound has been obtained on their quantum lifetimes \cite{BG.PRE.q}.
In the classical limit, the scattering is typically stochastic at each vertex of the graph so that the classical dynamics may be chaotic with a positive Kolmogorov-Sinai entropy per unit time on graphs with more than two vertices \cite{BG.PRE.cl}.

In this paper, we propose to study the integer quantum Hall effect (IQHE) on the basis of quantum graphs. This phenomenon which is manifested by the quantization of the Hall conductance is the object of many works since its discovery by von~Klitzing \cite{klitzing}. Thouless \emph{et al.} \cite{thoul} showed the important link between the Hall conductance and the energy spectra of independent electron's model, a result which led Osadchy and Avron \cite{osa}  to draw the phase diagram for Hofstadter's model \cite{hof}. This phase diagram is fractal and depicts infinitely many phases, each one characterized by the integer value of the Hall conductance. Such fractal phase diagrams have also been studied in cold atomic systems submitted to artificial gauge fields \cite{Goldman2007} and in models featuring continuous potentials which exhibit classical chaos \cite{PG93,SKG97}.

In this context, quantum graphs are interesting because the propagative medium is formed of one-dimensional continuous bonds, in contrast to the fully discrete lattices of the usual tight-binding models.  Yet, quantum graphs are simple enough that they can be investigated by analytical developments, as we shall show.

We first  study the energy spectra for infinite rectangular graphs perturbed by a strong magnetic field. We obtain the spectral properties of two-dimensional ($2D$) and three-dimensional ($3D$) systems.  This allows us to compare these systems from the viewpoint of chaology. We show that the eigenvalue equation of quantum graphs can be mapped onto the generalized Harper equation in the case of $2D$ rectangular lattices. We then compute the transverse conductivity of the $2D$ system with Kubo's formula which describes the linear response of this system to an external electric field, and obtain the quantization law through topological arguments. We then show the link between the Hall conductivity and the energy spectra by computing this quantity for Fermi energies located inside the many gaps of the spectra. We conclude this work with the presentation of a fractal phase diagram which describes the IQHE on $2D$ as well as $3D$ quantum graphs.

The paper is organized as follows.  The spectral properties of $2D$ and $3D$ quantum graphs are respectively presented in Secs. \ref{2D} and \ref{3D}. The Hall conductivity on quantum graphs is studied in Sec. \ref{Hall}.  The conclusions are drawn in Sec. \ref{conclusions}.

\section{Two-dimensional quantum graphs}
\label{2D}

\subsection{The eigenvalue equation}

A quantum graph is a set of vertices connected by unidimensional bonds on which a quantum wave propagates. Each bond of a rectangular lattice is characterized by a vertex coordinate  $(n,m)$ and by a direction  $b= 1,2$, and will be labeled $(n,m,b)$. Schrödinger's equation is satisfied on each bond
\begin{equation}
\left(- i \frac{d}{dr_b} - A^{(n,m,b)}\right)^2 \psi ^{(n,m,b)} (r_b)=k^2 \psi ^{(n,m,b)} (r_b)
\label{shsh}
\end{equation}
where $A^{(n,m,b)}$  is the potentiel vector component  along the bond $b$, $k=\sqrt{2 M E}/\hbar$ is the wave number, $M$ and $E$ are respectively the mass and the energy of the particle. In the following, we use units where $M=1$ and $\hbar=1$, except otherwise stated.  We consider an infinite graph forming a rectangular lattice submitted to a magnetic field $\boldsymbol{B}=\boldsymbol{\nabla} \times \boldsymbol{A}$ and we work in the Lorentz gauge $\boldsymbol{A}=(0,B \, x , 0)$, which maintains a discrete translational invariance along the $y$ direction. We suppose that the different bonds are directed along the $x$ (resp. $y$) axis with constant length $l_x$ (resp. $l_y$). The wave function $\psi ^{(n,m,1)} (x)$ (resp. $\psi ^{(n,m,2)} (y)$) is defined between vertex $(n,m)$ and $(n+1,m)$ [resp. $(n,m)$ and $(n,m+1)$]. Accordingly, we have that $x \in [0;l_x]$ and $y \in [0;l_y]$.

The solutions of Schrödinger's equation \eqref{shsh} are written as
\begin{align}
&\psi ^{(n,m,1)} (x)= C_{1}^{(n,m,1)} e ^{i k x} + C_{2}^{(n,m,1)} e ^{- i k x} \notag \\
&\psi ^{(n,m,2)} (y)= C_{1}^{(n,m,2)} e ^{i k y + i B n l_x y} + C_{2}^{(n,m,2)} e ^{- i k y  + i B n l_x y}
\label{grcoef}
\end{align}

We consider an anisotropic scattering model which results from the modification of the usual boundary conditions imposed at each vertex. We introduce two parameters $s$ and $r$ in these conditions
%\begin{widetext}
\begin{align}
\psi ^{(n,m,1)} (x=0)&=\psi ^{(n-1,m,1)} (x=l_x) \notag \\
&= s \, \psi ^{(n,m,2)} (y=0) \notag \\
&= s \, \psi ^{(n,m-1,2)} (y = l_y)  \notag \\
\label{conser}
\end{align}
\begin{align}
& r \left[ \frac{d\psi ^{(n,m,1)}}{dx}(x=0)  \,  
- \frac{d\psi ^{(n-1,m,1)}}{dx}(x=l_x) \, \right]  \notag \\
&
+ \frac{d\psi ^{(n,m,2)}}{dy}(y=0) - \frac{d\psi ^{(n,m-1,2)}}{dy}(y=l_y) \notag \\
&- i \, B \, n \, l_x \left[ \psi^{(n,m,2)}(y=0) - \psi^{(n,m-1,2)}(y=l_y) \right] =0 
\label{neuneuscat}
\end{align}
%\end{widetext}
where the particular case $r=s=1$ corresponds to an isotropic graph. One has to impose the unitarity of the scattering matrix $\hat S$. This matrix which plays an important role in the theory of quantum graphs  links the in-components $\chi ^{\, \textrm{in}}$ to the out-components $\chi ^{\, \textrm{out}}$ at each vertex. These components are written as follows for our model
\begin{align}
 &\chi ^{\, \textrm{out}}=\left( \begin{array}{ccc} C_{1}^{(n,m,1)} \\
C_{1}^{(n,m,2)} \\
C_{2}^{(n-1,m,1)} e ^{-i k l_x} \\
C_{2}^{(n,m-1,2)} e ^{- i k l_y  + i B n l_x l_y}
\end{array} \right) &\chi ^{\, \textrm{in}}= \left( \begin{array}{ccc} C_{2}^{(n,m,1)} \\
C_{2}^{(n,m,2)} \\
C_{1}^{(n-1,m,1)} e ^{i k l_x}\\
C_{1}^{(n,m-1,2)} e ^{i k l_y  + i B n l_x l_y}
\end{array} \right) 
\end{align}
where $C_{1,2}^{(n,m,b)}$ are the coefficients of the functions $\psi ^{(n,m,b)}$ as defined in Eqs. \eqref{grcoef}. The scattering matrix is then given by
\begin{equation}
\hat S= \frac{1}{rs+1} \left( \begin{array}{cccc} -1 & s & rs & s\\
r & -rs & r &1\\
rs & s & -1 & s \\
r & 1 & r & -rs  \end{array} \right)
\end{equation}
and the unitarity condition $\hat S^{\dagger} \hat S= \hat S \hat S^{\dagger}=\hat I$ sets $r=s$.
The unitarity condition guarantees the conservation of probability at each vertex.  This latter is expressed as
\begin{equation}
j^{(n-1,m,1)}+j^{(n,m-1,2)}=j^{(n,m,1)}+j^{(n,m,2)}
\end{equation}
with the probability current densities:
\begin{equation}
j^{(n,m,b)}=  \frac{1}{2 i} \left( \psi ^{(n,m,b)*} \frac{d  \psi ^{(n,m,b)}}{dr_b} - \frac{d  \psi ^{(n,m,b)*}}{dr_b}  \psi ^{(n,m,b)} \right) - A^{(n,m,b)} \vert \psi ^{(n,m,b)} \vert ^2 
\end{equation}

One can then write the solutions \eqref{grcoef} in the following form
%\begin{widetext}
\begin{align}
&\psi^{(n,m,1)} (x)= f_{n,m} \frac{\sin k(l_x - x)}{\sin k l_x} + f_{n+1,m} \frac{\sin k x}{\sin k l_x} \notag \\
&\psi ^{(n,m,2)} (y)= \frac{1}{r} \, f_{n,m} e ^{i B n l_x y}\frac{\sin k( l_y - y)}{\sin k l_y} +  \frac{1}{r} \, f_{n,m+1} e ^{i B n l_x (y - l_y)}\frac{\sin k y}{\sin k l_y } 
\label{psi-f}
\end{align}
%\end{widetext}
where the coefficients $f_{n,m}$ are defined by
%\begin{widetext}
\begin{align}
&f_{n,m}= \psi ^{(n,m,1)} (x=0)= r \, \psi ^{(n,m,2)} (y=0)=C_{1}^{(n,m,1)} + C_{2}^{(n,m,1)}=r \left( C_{1}^{(n,m,2)} + C_{2}^{(n,m,2)}\right) \notag \\
&f_{n+1,m}= \psi ^{(n,m,1)} (x=l_x)=C_{1}^{(n,m,1)} e ^{i k l_x } + C_{2}^{(n,m,1)} e ^{- i k l_x} \notag \\
&f_{n,m+1}= r \,  \psi ^{(n,m,2)} (y=l_y)= r \left( C_{1}^{(n,m,2)} e ^{i k l_y + i 2 \pi \Phi n} + C_{2}^{(n,m,2)} e ^{- i k l_y + i 2 \pi \Phi n}  \right)
\end{align}
%\end{widetext}
Here, $\Phi = B l_x l_y/ (2 \pi)$ is the magnetic flux through a unit cell and the condition \eqref{conser} has been applied.

The system being invariant under discrete translations along the $y$ axis, we are led to consider new wave functions $u^{(n,b)}$ defined in terms of the solutions $\psi ^{(n,m,b)}$ of Eq. \eqref{shsh} according to
\begin{align}
u^{(n,1)}(x)& \equiv \psi^{(n,m,1)} (x) \, e ^{- i k _x n l_x - i k_y m l_y} \label{u1} \\
u^{(n,2)}(y)& \equiv \psi^{(n,m,2)} (y) \, e ^{- i k _x n l_x - i k_y  m l_y} \label{u2}
\end{align}
Similarly, the coefficients $g_n$ are defined as
\begin{align}
g_n& \equiv f_{n,m} \, e ^{- i k _x n l_x - i k_y m l_y} 
\label{g}
\end{align}

According to Eqs. \eqref{psi-f}, the new wave functions $u^{(n,b)}$ are
written in terms of new coefficients \eqref{g} as
\begin{widetext}
\begin{align}
&u^{(n,1)} (x) = g_{n}  \frac{\sin k (l_x -  x)}{\sin k l_x}  + g_{n+1} \frac{\sin k x}{\sin k l_x} e^{i k_x l_x} \\
&u^{(n,2)} (y) =  \frac{1}{r} \, g_{n} \left[ e ^{i B n l_x y}\frac{\sin k (l_y - y)}{\sin k l_y} + e ^{i B n l_x (y - l_y)}\frac{\sin k y}{\sin k l_y} e^{i k_y l_y} \right]
\end{align}
\end{widetext}

The probability conservation \eqref{neuneuscat} implies
\begin{align}
& e ^{i k _x l_x} \, g_{n+1} + e ^{-i k _x l_x} \, g_{n-1}  + \frac{2}{r^2} \, \frac{\sin k l_x}{\sin kl_y} \cos (2 \pi \Phi n - k _{y} l_y) \, g_n \notag \\
&= 2 \left(\cos k l_x + \frac{\cos k l_y}{r^2} \,  \frac{\sin k l_x}{\sin k l_y} \right)  \, g_n
\label{eigenvalue}
\end{align}

If we set
\begin{align}
&\Lambda =\frac{1}{r^2}\, \frac{\sin kl_x}{\sin kl_y} \label{Lambda} \\
&\mathcal{E}= 2 \, \left(\cos k l_x + \frac{\cos k l_y}{r^2} \,  \frac{\sin k l_x}{\sin k l_y} \right) \label{Energy}
\end{align}
one finds the generalized Harper equation
\begin{equation}
e ^{i k _x l_x} \, g_{n+1} + e ^{-i k _x l_x} \, g_{n-1}  + 2\Lambda \cos (2 \pi \Phi n - k _{y} l_y) \, g_n = \mathcal{E}  \, g_n
\label{superharp}
\end{equation}
which reduces to Harper's equation for Hofstadter's model when $l_x=l_y$ and $r=1$. This shows that the problem of the quantum graph can be mapped onto a similar problem for the anisotropic tight-binding Hamiltonian with the transfer coefficients $t_a$ and $t_b$ \cite{mont}.  The correspondence is established with the anisotropy ratio $\Lambda=t_b/t_a$.  However, an important difference is that the parameter $\Lambda$ and the energy $\mathcal{E}$ are independent in the tight-binding model, albeit they both depend on the wave number $k=\sqrt{2E}$ in the quantum graph.  In this regard, the eigenvalue problem is more complicated in quantum graphs.

We notice that we would have obtained the dual generalized Harper equation
\begin{equation}
e ^{i k _y l_y} \, h_{m+1} + e ^{-i k _y l_y} \, h_{m-1}  + \frac{2}{\Lambda} \cos (2 \pi \Phi m + k _{x} l_x) \, h_m = \frac{\mathcal{E}}{\Lambda}  \, h_m
\label{superharp.m}
\end{equation}
if we had used another gauge with the vector potential $\boldsymbol{A}'=(-B \, y, 0 , 0)$. 
In principle, the eigenvalue equations (\ref{superharp}) and (\ref{superharp.m}) should lead to the same energy spectrum since they correspond to the same magnetic field $\boldsymbol{B}=\boldsymbol{\nabla} \times \boldsymbol{A}'=\boldsymbol{\nabla} \times \boldsymbol{A}$.  Indeed, Eq. (\ref{superharp.m}) is derived from Eq. (\ref{superharp}) thanks to the duality transformation \cite{aubry}:
\begin{equation}
h_m \equiv \sum_{n=-\infty}^{+\infty} g_n \; e^{-i2\pi\Phi mn}
\end{equation}

Our aim is to obtain the spectra in the plane of the magnetic flux $\Phi$ versus the wave number $k=\sqrt{2E}$. This latter is directly related to the energy $E=k^2/2$ and has the advantage to be well adapted to quantum graphs since the spectra tend to distribute themselves uniformly along the wave number $k$-axis, which is not the case along the energy axis.  The method to obtain the spectra is to consider the rational values of the magnetic flux $\Phi = p/q$, with $p,q \in \mathbb{Z}$.  The rational numbers are dense in the real numbers so that we may hope to display the structure of the spectrum by plotting the spectra for all the rational numbers $\Phi = p/q$ up to a large enough integer $q$.

If the magnetic flux is rational $\Phi = p/q$, we may assume that the functions $u^{(n,b)}$ satisfy the periodic boundary conditions
\begin{align}
&u^{(n+q,1)}(x)=u^{(n,1)}(x)   \notag \\
&u^{(n+q,2)}(y)=u^{(n,2)}(y) \; e^{i 2 \pi p \frac{y}{l_y}}  
\label{cbp}
\end{align}
In this case, the Bloch parameters take their values in the ranges $k_x \in [- \frac{\pi} {q l_x} , \frac{\pi} {q l_x}]$ and $k_y \in [- \frac{\pi}{l_y} , \frac{\pi}{ l_y} ]$ which delimit the first Brillouin zone. 

As a consequence of Eqs. \eqref{u1}, \eqref{u2}, and \eqref{cbp}, the coefficients \eqref{g} satisfy the periodic condition
\begin{align}
g_{n+q}=g_{n}
\end{align}
and the eigenvalue equation (\ref{superharp}) can be solved by requiring that the corresponding characteristic $q\times q$ determinant is vanishing.

\subsection{Quantum graphs without magnetic field}

If the magnetic flux vanishes $\Phi=0$, the eigenvalue equation (\ref{eigenvalue}) can be solved by taking $g_{n+1}=g_n$.  The eigenvalues are thus given by the zeros of the following function of the wave number $k$:
\begin{equation}
f(k)=r^2 \left(\cos kl_x - \cos k_xl_x \right) \sin kl_y + \left(\cos kl_y - \cos k_yl_y \right) \sin kl_x = 0
\label{f(k).2D}
\end{equation}
For fixed values of Bloch's parameters $k_x$ and $k_y$, the spectrum is discrete.  A continuous band spectrum is obtained by varying Bloch's parameters in the first Brillouin zone delimited by $k_x \in [- \frac{\pi} {l_x} , \frac{\pi} {l_x}]$ and $k_y \in [- \frac{\pi}{l_y} , \frac{\pi}{ l_y} ]$. An example of band spectrum is depicted in Fig. \ref{2Dband.fig}.

\begin{widetext}
\begin{center} 
\begin{figure}
\begin{center} 
{\scalebox{2.9}{\includegraphics{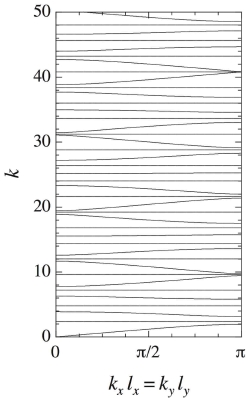}}} 
\caption{\label{2Dband.fig} Band spectrum of the $2D$ graph without magnetic field $\Phi=0$.  The bond lengths are $l_x=1$ and $l_y=(1+\sqrt{5})/2$.  The bands are depicted in the plane of the wave number $k=\sqrt{2E}$ versus the Bloch parameters $k_xl_x=k_yl_y$.} 
\end{center}
\end{figure} 
\end{center} 
\end{widetext}

For fixed values of the Bloch parameters, the spectrum is discrete as aforementioned.  In this case, we can study
the statistics of the level spacings: 
\begin{equation}
S=\frac{k_{i+1}-k_i}{\langle k_{i+1}-k_i\rangle}
\label{spacing}
\end{equation}
where $k_i$ with $i\in\mathbb{N}$ are the roots of Eq. (\ref{f(k).2D}): $f(k_i)=0$.
We observe in Fig. \ref{2Dspacing.fig} that the level spacing distribution is empty around zero spacing for $k_xl_x=1$ and $k_yl_y=0.7$.  This gap is due to the fact that there are only two incommensurate lengths $l_x$ and $l_y$ in the $2D$ graph. As a consequence, it is known that the spacing distribution should generically present such a gap \cite{BG.JSP}.

\begin{widetext}
\begin{center} 
\begin{figure}
\begin{center} 
{\scalebox{1.5}{\includegraphics{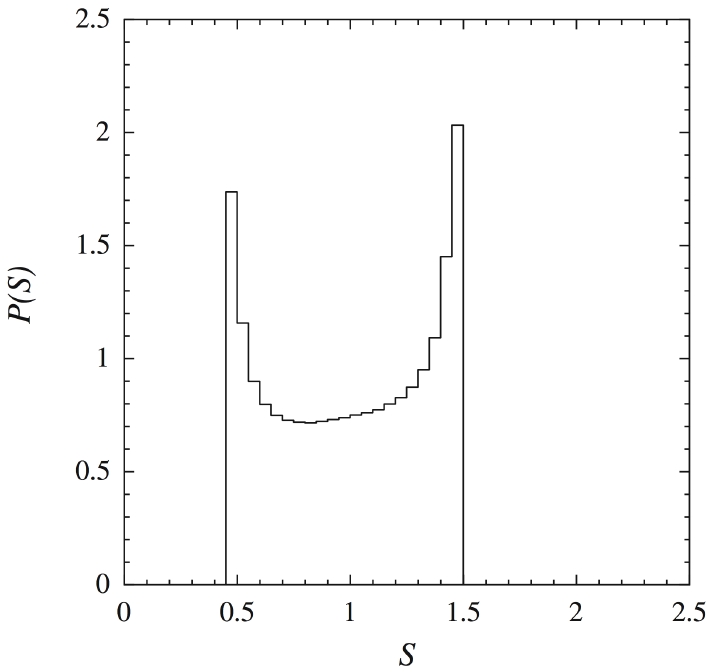}}} 
\caption{\label{2Dspacing.fig} The level spacing probability density $P(S)$ for the $2D$ graph without magnetic field $\Phi=0$ at the values $k_xl_x=1$ and $k_yl_y=0.7$ of Bloch parameters.  The bond lengths are $l_x=1$ and $l_y=(1+\sqrt{5})/2$.  We observe the expected gap around zero spacing.} 
\end{center}
\end{figure} 
\end{center} 
\end{widetext}

In the classical limit, point particles move with the velocity $v$ on the one-dimensional bonds and are scattered stochastically at each vertex \cite{BG.PRE.cl}.  In the case where $r=1$, the probability to be scattered in one of the four directions of the lattice are equal to $1/4$.  Accordingly, the particle undergoes a diffusive random walk on the lattice, the properties of which can be calculated with the methods of Ref. \cite{BG.PRE.cl}.  The diffusion coefficients take the values $D_x = \frac{vl_x^2}{2(l_x+l_y)}$ and $D_y = \frac{vl_y^2}{2(l_x+l_y)}$.  The Kolmogorov-Sinai entropy per unit time is equal to $h_{\rm KS} = \frac{v\ln 4}{l_x+l_y}$ and its positivity shows that the classical motion is chaotic.

\subsection{Quantum graphs with magnetic field}

\begin{center} 
\begin{figure}%[!h] 
\begin{center} 
\vspace{-0cm}
\hspace*{-0cm} 
{\scalebox{4.9}{\includegraphics{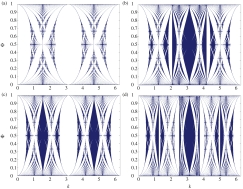}}}
\caption{\label{graphin} Spectra of the quantum graphs for the cases: (a) $l_x=1$, (b) $l_x=1.5$, (c) $l_x=2$, and (d) $l_x=2.5$ with $l_y=1$ and $r=1$.  The wave number $k=\sqrt{2 E}$ is along the horizontal axis and the magnetic flux $\Phi$ along the vertical axis.} \end{center} 
\end{figure} 
\end{center}

We solve numerically the generalized Harper equation in order to obtain the energy spectrum as a function of the magnetic flux. Several spectra are shown in Figs. \ref{graphin}(a)-(d) in terms of the wave number $k=\sqrt{2 E}$ for the cases $l_x=1, 1.5, 2, 2.5$ with $l_y=1$ and $r=1$.

In the case $l_x=l_y=1$ and $r=1$, the lattice has the square symmetry and the spectrum in Fig. \ref{graphin}(a) resembles the Hofstadter butterfly \cite{hof}. We can show that it is identical to the Hofstadter butterfly up to a deformation.
Indeed, Eqs. (\ref{Lambda}) and (\ref{Energy}) give $\Lambda=1$ and $\mathcal{E}=4\cos k$ in this case.
Accordingly, Eq. (\ref{superharp}) reduces to Harper's equation which leads to the Hofstadter butterfly represented in the plane of the magnetic flux $\Phi$ versus the energy $\mathcal{E}$.  Figure \ref{graphin}(a) depicts the butterfly versus the wave number $k={\rm arccos}(\mathcal{E}/4)$.  The butterfly is thus only deformed by this change of variable.

However, the anisotropy enhanced by the difference $l_x \ne l_y$ closes many gaps [dark zones in Figs. \ref{graphin}(b)-(d)], which reminds a similar phenomenon observed in the anisotropic Hofstadter model \cite{mont}. The spectrum of the latter model is described by the Harper equation \eqref{superharp} and is darkened when the anisotropy ratio $\Lambda =\frac{t_b}{t_a}$ differs from unity. We have a similar phenomenon in the spectrum of quantum graphs although it is complicated by the common dependence of $\Lambda$ and $\mathcal{E}$ on the wave number $k$.  We observe that the dark zones appear away from the values $k_{s}$ of the wave number for which the anisotropy coefficient is close to unity: $\Lambda = \vert \frac{\sin k_{s} l_x}{r^2 \sin k_{s} l_y}  \vert \simeq 1$.  The reasons are that the generalized Harper equations (\ref{superharp}) and (\ref{superharp.m}) are known to admit extended eigenstates if $\Lambda$ is different from unity \cite{aubry}, and that continuous spectra are typically associated with extended eigenstates.   In contrast, the spectrum displays fractal structures reminiscent of  the Hofstadter butterfly around the special values $k_{s}$ where the anisotropy ratio approaches unity.

In the case $l_x=1.5$ depicted in Fig. \ref{graphin}(b), the anisotropy ratio is unity if $\vert \sin k_s\vert = \vert \sin(3k_s/2)\vert$.  This condition is satisfied at $k_s=\frac{2\pi}{5}, \frac{4\pi}{5}, \frac{6\pi}{5}, \frac{8\pi}{5}, ...$
We clearly see in Fig. \ref{graphin}(b) that the spectrum displays the fractal structures of  the Hofstadter butterfly around these values, while it darkens away.

Similarly, we see in the case $l_x=2$ depicted in Fig. \ref{graphin}(c) that the spectrum looks locally as the Hofstadter butterfly around the special values $k_s=\frac{\pi}{3}, \frac{2\pi}{3}, \frac{4\pi}{3}, \frac{5\pi}{3}, ...$ where $\Lambda=1$. 

In the case $l_x=2.5$ depicted in Fig. \ref{graphin}(d), this also happens around the special values $k_s=0.898,1.795,2.094,2.693,3.590,4.189,4.488,5.386,...$ where $\Lambda=1$.  Here, the situation is more subtle because, for instance, the values $k_s=1.795$ and $k_s=2.094$ are very close to each other and the anisotropy ratio remains close to unity between these values. This explains that the spectrum looks as a Hofstadter butterfly only once in this interval and similarly for $4.189<k_s<4.488$.

We notice that the spectra are periodic in the wave number if the two lengths $l_x$ and $l_y$ are commensurate.  They form non-periodic structures as the wave number increases if the lengths are incommensurate. We finally note that for $l_x=l_y=n$, where $n$ is an integer, the spectrum represents exactly $n$ Hofstadter butterflies for $k \in [0, \pi ]$. In the anisotropic situations depicted in Fig. \ref{graphin}(c) in which $l_x=n$ and $l_y=1$, we find exactly $n$ butterfly-like structures for $k \in [0, \pi ]$.  

\section{Three-dimensional quantum graphs}
\label{3D}

\subsection{The eigenvalue equation}

The insertion of a third dimension in quantum graphs submitted to magnetic fields is expected to yield drastic changes in the spectral properties of these systems.  Recent works have been investigating the effects due to three-dimensionality in the Hofstadter model \cite{koshinopb,koshinoprl,koshinoprb}. These studies led to the conclusion that Hofstadter butterfly-like fractal patterns still exist in $3D$ space systems under certain conditions. 

We here show that fractal patterns also persist in $3D$ quantum graphs. Each bond of the $3D$ lattice is characterized by a vertex coordinate  $(n,m,l)$ and by a direction  $b= 1,2,3$, and will be labeled $(n,m,l,b)$. We suppose that the magnetic field is still oriented along the $z$-axis and that we work with the Lorentz gauge where $\boldsymbol{A}=(0,B \, x , 0)$, so that the wave function along the bond $(n,m,l,3)$ is a solution of the Schrödinger equation
\begin{equation}
\left(- i \frac{d}{d z}\right)^2 \psi ^{(n,m,l,3)} (z)=k^2 \psi ^{(n,m,l,3)} (z)
\end{equation}

We can then treat this system using the method described in the previous Sec. \ref{2D}. We impose boundary conditions similar to the $2D$ case with
\begin{align}
&  \frac{d\psi ^{(n,m,l,1)}}{dx}(x=0) \,  
- \frac{d\psi ^{(n-1,m,l,1)}}{dx}(x=l_x) 
+ \frac{d\psi ^{(n,m,l,2)}}{dy}(y=0) - \frac{d\psi ^{(n,m-1,l,2)}}{dy}(y=l_y) \notag \\
&
 - i \, B \, n l_x \left[ \psi ^{(n,m,l,2)}(y=0) - \psi^{(n,m-1,l,2)}(y=l_y) \right] \notag \\
&
+ \gamma \, \left[\frac{d\psi ^{(n,m,l,3)}}{dz}(z=0) \,  
- \frac{d\psi ^{(n,m,l-1,3)}}{dz}(z=l_z) \right] =0 \label{neuneuscat2}
\end{align}
where the parameter $\gamma \in [0,1]$ maps the $2D$ system ($\gamma =0$) to the $3D$ isotropic system ($\gamma=1$). 

The wave function is explicitly written as
\begin{equation}
\psi^{(n,m,l,3)} (z)= f_{n,m,l} \frac{\sin k( l_z - z)}{\sin k l_z} + f_{n,m,l+1}  \frac{\sin k z}{\sin k l_z}
\end{equation}
Setting 
\begin{align}
g_n& \equiv f_{n,m,l} \, e ^{- i k _x n l_x - i k_y m l_y - i k_z l l_z} 
\end{align}
with $k_x \in [- \frac{ \pi}{q l_x}, \frac{ \pi}{q l_x}]$, $k_y \in [- \frac{ \pi}{l_y}, \frac{ \pi}{l_y}]$ and $k_z \in [- \frac{ \pi}{l_z}, \frac{ \pi}{l_z}]$, one finds the generalized $3D$ Harper equation
\begin{widetext}
\begin{equation}
e ^{i k _x l_x} \, g_{n+1} + e ^{ - i k _x l_x} \, g_{n-1}+  \left[ 2 \, \Lambda \, \cos(2 \pi \Phi n - k _y l_y) + 2  \,  \tilde{\Lambda} \, \cos k _z l_z \right] \, g_n= \mathcal{E} \, g_n   
\label{3Dsuperharp}
\end{equation}
\end{widetext}
with the coefficients
\begin{eqnarray}
\Lambda &=& \frac{\sin k l_x}{\sin k l_y} \label{3DL}\\
\tilde{\Lambda } &=& \gamma^2 \frac{\sin k l_x}{\sin k l_z} \label{3DL'}\\
\mathcal{E} &=& 2 \left( {\rm cotg}\, k l_x + {\rm cotg}\, k l_y + \gamma^2 {\rm cotg}\, k l_z \right) \sin k l_x \label{3DE}
\end{eqnarray}
As in the $2D$ system, these coefficients all depend on the wave number $k$.

\subsection{Quantum graphs without magnetic field}

If the magnetic flux vanishes $\Phi=0$, we may assume that the solution is periodic with $g_{n+1}=g_n$, in which case the eigenvalue equation (\ref{3Dsuperharp}) with the coefficients (\ref{3DL})-(\ref{3DE}) becomes
\begin{equation}
f(k)=\left(\cos kl_x - \cos k_xl_x \right) \sin kl_y \sin kl_z + \left(\cos kl_y - \cos k_yl_y \right) \sin kl_x  \sin kl_z + \gamma^2 \left(\cos kl_z - \cos k_zl_z \right) \sin kl_x  \sin kl_y = 0
\end{equation}
Because of its spatial periodicity, the system has continuous band spectra, an example of which is depicted in Fig. \ref{3Dband.fig}.  Contrary to the $2D$ system, Wigner repulsion manifests itself in the level spacing statistics.   The Wigner repulsion has for consequence that a typical level spacing probability density behaves as $P(S)\sim S$ for $S\to 0$, as observed in Fig. \ref{3Dspacing.fig}.  Consequently, there is no gap at small level spacings contrary to what happens in Fig. \ref{2Dspacing.fig} for the $2D$ case.  The reason is that the $3D$ lattices typically have three incommensurate lengths, which is the minimum number for Wigner repulsion to manifest itself \cite{BG.JSP}.

\begin{widetext}
\begin{center} 
\begin{figure}
\begin{center} 
{\scalebox{3}{\includegraphics{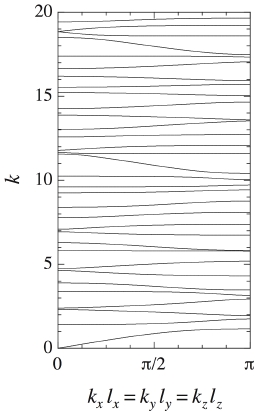}}} 
\caption{\label{3Dband.fig} Band spectrum of the $3D$ graph without magnetic field $\Phi=0$.  The bond lengths are $l_x=1$, $l_y=(1+\sqrt{5})/2$, and $l_z=\exp(1)$.  The bands are depicted in the plane of the wave number $k=\sqrt{2E}$ versus the Bloch parameters $k_xl_x=k_yl_y=k_zl_z$.} 
\end{center}
\end{figure} 
\end{center} 
\end{widetext}

\begin{widetext}
\begin{center} 
\begin{figure}
\begin{center} 
{\scalebox{2}{\includegraphics{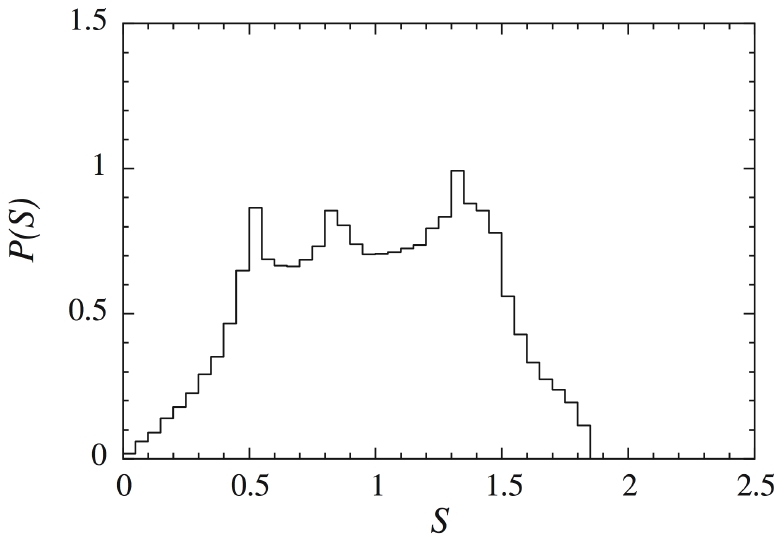}}} 
\caption{\label{3Dspacing.fig} The level spacing probability density $P(S)$ for the $3D$ graph without magnetic field $\Phi=0$ at the values $k_xl_x=1.5$, $k_yl_y=1$ and $k_zl_z=0.7$ of Bloch parameters.  The bond lengths are $l_x=1$, $l_y=(1+\sqrt{5})/2$, and $l_z=\exp(1)$.  We observe the behavior characteristic of Wigner repulsion: $P(S) \sim S$ for $S\to 0$.} 
\end{center}
\end{figure} 
\end{center} 
\end{widetext}

\subsection{Quantum graphs with magnetic field}

\begin{center} 
\begin{figure}%[!h] 
\begin{center} 
{\scalebox{3.95}{\includegraphics{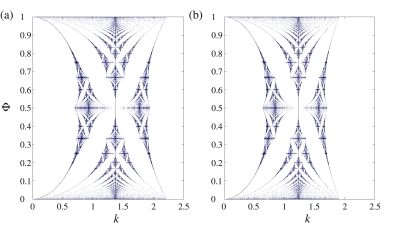}}}
\caption{\label{graphin3D} Spectra of the quantum graphs for: (a) $\gamma=0.5$, (b) $\gamma=1$, with $l_x=l_y=l_z=1$ and $k_z=0$.  The wave number $k=\sqrt{2 E}$ is along the horizontal axis and the magnetic flux $\Phi$ along the vertical axis.} \end{center} 
\end{figure} 
\end{center}

\begin{center} 
\begin{figure}%[!h] 
\begin{center} 
\begin{tabular}{c@{\hspace{0cm}}c} 
\vspace{-0cm}
\hspace*{-0cm} 
{\scalebox{2.95}{\includegraphics{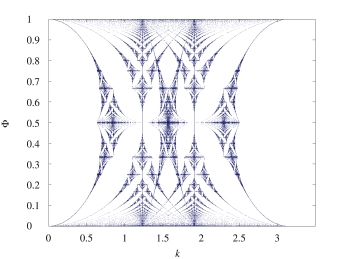}}}
\vspace{-00pt}
\end{tabular}
\caption{\label{graphin3D2} Spectra of the quantum graphs for $\gamma=1$, with $l_x=l_y=l_z=1$ and $k_z=n \pi$, $n \in \mathbb{Z}$.  The wave number $k=\sqrt{2 E}$ is along the horizontal axis and the magnetic flux $\Phi$ along the vertical axis.} \end{center} 
\end{figure} 
\end{center}

We solved the generalized $3D$ Harper equation (\ref{3Dsuperharp}) with $\Phi\neq 0$ in order to find the spectrum associated to the system for different values of $\gamma$. As shown in Fig. \ref{graphin3D}, the butterfly seems to lose its initial symmetric shape as $\gamma$ differs from zero, and eventually forms a new fractal structure for $\gamma=1$. It is worth noticing that the many gaps which compose the Hofstadter butterfly are conserved as one maps the $2D$ system to the $3D$ system.  Figure \ref{graphin3D} depicts the spectra corresponding to a single value of the Bloch parameter $k_z$, which corresponds to the situation of a very flattened $3D$ system. As the graph thickens, $k_z$ takes more values between $- \pi/l_z$ and $\pi/l_z$ and the spectrum appears as the superposition of several deformed butterflies. As the three-dimensionality becomes important, gaps close and the spectrum darkens as shown in Fig. \ref{graphin3D2}.

\section{Quantum Hall effect on quantum graphs}
\label{Hall}

\subsection{Kubo formula on quantum graphs}

In order to study the integer quantum Hall effect, we consider independent Fermions on a $2D$ quantum graph and evaluate the antisymmetric component $\sigma _{xy}=-\sigma _{yx}$ of the conductivity tensor with Kubo's formula for zero temperature. The latter relates $\sigma _{xy}$ to the current intensity $\hat{\boldsymbol{J}}=(\hat J_x,\hat J_y)$ and can be written in the following way
\begin{widetext}
\begin{equation}
\sigma _{xy}=\frac{1}{i V} \sum_{\bf N} \frac{\langle {\bf N} \vert \hat J_x \vert 0 \rangle \langle 0 \vert \hat J_y \vert {\bf N} \rangle - (x\leftrightarrow y)}{(E_0- E_{\bf N})^2} 
\label{condu2}
\end{equation}
\end{widetext}
where $\vert {\bf N}\rangle$ denotes the eigenstates of the many-particle
Hamiltonian $\hat{\mathcal{H}}$ of eigenvalue $E_{\bf N}$, $\vert 0 \rangle$ is the ground state, and $V=\sum_{(n,m,b)} l_b$ is the system ``volume" given by the total lengths of the bonds in a unit cell of the periodic lattice.  $l_b$ denotes the length of the bond $b$.

We define the current density which circulates along the bond $(n,m,b)$ in the direction $x$ by
\begin{equation}
\hat{j}_x^{(n,m,b)}=  \frac{1}{2 i} \left( \hat\Psi ^{(n,m,b)\dagger} \frac{d  \hat\Psi ^{(n,m,b)}}{d x} - \frac{d  \hat\Psi ^{(n,m,b)\dagger}}{d x}  \hat\Psi ^{(n,m,i)} \right) - A_x^{(n,m,b)} \hat\Psi ^{(n,m,b)\dagger} \hat\Psi^{(n,m,b)}
\label{current_density}
\end{equation}
where $A_x^{(n,m,b)}$ is the $x$-component of the potentiel vector along the bond $(n,m,b)$ and $\hat\Psi^{(n,m,b)}$ is the field operator defined on the bond. In the second quantization formalism, the field operator and the adjoint field operator are respectively given by
\begin{align}
\hat\Psi ^{(n,m,b)}&=\sum_{\nu} \psi_{\nu}^{(n,m,b)} \, \hat a_{\nu} \label{Psi}\\
\hat\Psi ^{(n,m,b)\dagger } &=\sum_{\nu'} \psi^{(n,m,b)*} _{\nu' } \, \hat a^{\dagger }_{\nu'} \label{Psi_dagger}
\end{align}
where $\hat a_{\nu}$ and $\hat a_{\nu}^{\dagger}$ are the annihilation and creation operators satisfying the anticommutation relation: $\hat a_{\nu}\hat a_{\nu'}^{\dagger}+\hat a_{\nu'}^{\dagger}\hat a_{\nu}=\delta_{\nu\nu'}$.
If the system is in the Fock state $\vert{\bf N}\rangle = \vert N_1 N_2 N_3\cdots N_{\nu}\cdots\rangle$ with $N_{\nu}=0$ or $1$ Fermion on each single-particle wave function $\psi_{\nu}$, the total energy of the system takes the value
\begin{equation}
E_{\bf N}= \sum_{\nu} N_{\nu} \, \epsilon _{\nu} 
\end{equation}
where $\epsilon _{\nu}$ is the energy corresponding to the single-particle wave function $\psi_{\nu}$ and $N_{\nu}$ is the corresponding occupation number ($N_{\nu}=0,1$).

The current intensity which circulates in a unit cell of the lattice is defined by 
\begin{equation}
\hat J_x = \sum_{(n,m,b)} \int _0^{l_b} \hat j_x^{(n,m,b)} dr_b
\end{equation}
where the sum over $(n,m,b)$ extends over all the bonds of the unit cell.
Substituting the expression (\ref{current_density}) of the current density and using Eqs. (\ref{Psi})-(\ref{Psi_dagger}), we find that the current intensity is given by
\begin{align}
\hat J_x &= \sum_{(n,m,b)}\sum _{\nu , \nu'}   \int _0^{l_b} dr_b \left[ \frac{1}{2 i} \left( \psi ^{(n,m,b)*}_{\nu '} \frac{d \psi ^{(n,m,b)}_{\nu}}{d x} - \frac{d \psi ^{(n,m,b)*}_{\nu '}}{d x}   \psi ^{(n,m,b)}_{\nu } \right) - A_x^{(n,m,b)} \psi ^{(n,m,b) *}_{\nu '} \psi ^{(n,m,b)}_{\nu} \right] \hat a^{\dagger}_{\nu '} \, \hat a_{\nu} \\
&\equiv \sum _{\nu , \nu'}  \langle \nu'\vert \hat v_x\vert \nu\rangle \, \hat a^{\dagger}_{\nu '} \, \hat a_{\nu}
\label{courgraph}
\end{align}
The quantities $\langle \nu'\vert \hat v_x\vert \nu\rangle$ which are here introduced are the matrix elements of the particle velocity in the single-particle wave functions $\psi_{\nu}$. The scalar product in the space of the single-particle wave functions on the graph is defined by
\begin{equation}
\langle \phi \vert \psi \rangle = \sum_{(n,m,b)} \int_0^{l_b} \phi^{(n,m,b) *} \psi ^{(n,m,b)} dr_b
\label{scal}
\end{equation}
where the sum $\sum_{(n,m,b)}$ accounts for the contribution of all the bonds in the unit cell.

Since the operator $\hat J_x$ contains one annihilation and one creation operator, the matrix element $\langle {\bf N} \vert \hat J_x \vert 0 \rangle$ is non vanishing only for Fock states 
\begin{equation}
\vert{\bf N}\rangle = \vert 111\cdots 111\underbrace{0}_{\alpha}111\cdots 111_{{\displaystyle\uparrow}\atop{\displaystyle\epsilon _{\rm F}}} 000 \cdots 000\underbrace{1}_{\beta}000 \cdots \rangle
\label{nstate2}
\end{equation}
with one hole and one particle. $\epsilon_{\rm F}$ denotes the Fermi energy. Indeed, applying an annihilation operator followed by a creation operator on the ground state gives such a state up to a phase:
\begin{equation}
\hat a_{\beta}^{\dagger}\hat a_{\alpha} \vert 111\cdots 111_{{\displaystyle\uparrow}\atop{\displaystyle\epsilon _{\rm F}}} 000 \cdots 000 \cdots \rangle  = (-1)^{\alpha+N_{\rm F}}\vert 111\cdots 111\underbrace{0}_{\alpha}111\cdots 111_{{\displaystyle\uparrow}\atop{\displaystyle\epsilon _{\rm F}}} 000 \cdots 000\underbrace{1}_{\beta}000 \cdots \rangle
\end{equation}
where $N_{\rm F}$ is the total number of Fermions and $\alpha$ the integer labelling the corresponding single-particle state.  Therefore, the matrix element of the current intensity is given by
\begin{equation}
\langle {\bf N} \vert \hat J_x \vert 0 \rangle = (-1)^{\alpha+N_{\rm F}}
\langle \beta\vert \hat v_x\vert \alpha\rangle
\end{equation}
for the Fock state (\ref{nstate2}) and zero otherwise.  Moreover, we notice that the energy of this Fock state (\ref{nstate2}) is equal to
\begin{equation}
E_{\bf N} - E_0= \epsilon_{\beta}-\epsilon_{\alpha}
\label{energy_state2}
\end{equation}
The energy of the single-particle state $\vert\alpha\rangle$ is below the Fermi energy $\epsilon_{\alpha}< \epsilon_{\rm F}$, while the situation is opposite for the state $\vert\beta\rangle$: $\epsilon_{\beta}> \epsilon_{\rm F}$.

Therefore, the conductivity (\ref{condu2}) can be expressed in terms of single-particle wave functions of the quantum graph as
\begin{equation}
\sigma _{xy}=\frac{1}{i V} \sum_{\epsilon_{\alpha}< \epsilon_{\rm F}} \sum_{\epsilon_{\beta}>\epsilon_{\rm F}}  \frac{\langle \beta\vert \hat v_x\vert \alpha\rangle \langle \alpha\vert \hat v_y\vert \beta\rangle - (x\leftrightarrow y)}{(\epsilon_{\alpha}-\epsilon_{\beta})^2} 
\label{condu3}
\end{equation}

\subsection{Quantization of transverse conductivity}

We consider the single-particle wave functions \eqref{u1} and \eqref{u2} with the periodic boundary conditions \eqref{cbp}. The corresponding single-particle Hamiltonian is defined by
\begin{equation}
\hat H = \frac{1}{2} \left( -i \partial _{x} - A_x +  k_x\right)^2 + \frac{1}{2} \left( -i \partial _{y} - A_y +  k_y\right)^2
\label{gram}
\end{equation} 
where $k_x$ and $k_y$ are the components of the wave vector. The latter are Bloch parameters which take their values on the first Brillouin's zone of the reciprocal space: $k_x \in [- \frac{\pi} {q l_x} , \frac{\pi} {q l_x}]$ and $k_y \in [- \frac{\pi}{l_y} , \frac{\pi}{ l_y} ]$.  The velocity operator is given in terms of this Hamiltonian by
\begin{equation}
\hat v_x = \frac{\partial\hat H}{\partial k_x}
\label{velocity}
\end{equation} 

We consider the Hamiltonian eigenstates  $u_{\alpha}^{(n,b)}$ which satisfy Schrödinger's equation
\begin{equation}
\hat H u_{\alpha}^{(n,b)} = \epsilon _{\alpha} u_{\alpha}^{(n,b)}
\label{sheq}
\end{equation}
and the periodic boundary conditions \eqref{cbp}.  Differentiating the eigenvalue equation (\ref{sheq}) with respect to one component of the wave vector and taking the scalar product with another eigenstate, we get
\begin{equation}
\langle \beta\vert \hat v_x\vert \alpha\rangle \langle  = \langle \beta \vert \frac{\partial \hat H }{\partial k_{x}} \vert \alpha \rangle = (\epsilon _{\alpha} - \epsilon _{\beta}) \, \langle \beta \vert  \frac{ \partial  \alpha }{\partial k_{x}} \rangle
\label{hun}
\end{equation}
with the short notations $\vert\alpha\rangle \equiv  \vert u_{\alpha}\rangle$ and $\vert \partial_{k_x}  \alpha  \rangle \equiv \vert  \partial_{k_x}  u_{\alpha} \rangle$.

Using the relation $\langle\beta\vert\partial_{k_x}\alpha\rangle = - \langle\partial_{k_x}\beta\vert\alpha\rangle$ obtained by differentiating the orthonormality condition $\langle\alpha\vert\beta\rangle=\delta_{\alpha\beta}$ together with the completeness relation $\left(\sum_{\epsilon_{\alpha}<\epsilon_{\rm F}} + \sum_{\epsilon_{\alpha}>\epsilon_{\rm F}}\right) \vert\alpha\rangle\langle\alpha\vert = \hat I$, we find that the conductivity can be written as \cite{thoul}
\begin{equation}
\sigma _{xy}=\frac{1}{i V} \sum_{\epsilon_{\alpha}< \epsilon_{\rm F}}    \left\langle \frac{ \partial  \alpha }{\partial k_{y}} \vert \frac{ \partial  \alpha }{\partial k_{x}}\right\rangle - \left\langle \frac{ \partial  \alpha }{\partial k_{x}} \vert \frac{ \partial  \alpha }{\partial k_{y}}\right\rangle 
\label{condu4}
\end{equation}

If the Fermi level falls inside a spectral gap, the sum $\sum_{\epsilon_{\alpha}< \epsilon_{\rm F}}$ over all the states $\vert\alpha\rangle$ below the Fermi level can be decomposed into a sum over the occupied bands and a sum over the states $\vert\alpha\rangle$ inside a band.  This latter can be performed as an integral over the values of the Bloch parameters in the first Brillouin zone which forms a torus $\mathbb{T}^2$.  Accordingly, we have that
\begin{equation}
\sum_{\epsilon_{\alpha}< \epsilon_{\rm F}} = \sum_{\rm occ. \, bands} \, \,  \sum_{\epsilon_{\alpha} \,  \rm{ in \, occ. band}} = \sum_{\rm occ. \, bands} \frac{V}{(2\pi)2}\int_{\mathbb{T}^2} d \boldsymbol{k}
\end{equation}
Finally, using Stokes theorem in order to transform the two-dimensional integral over the torus into a line integral over the border of the torus, the transverse conductivity becomes
\begin{equation}
\sigma _{xy} = \frac{e^2}{2\pi i h} \sum_{\rm occ. \, bands}  \oint \left( \langle \alpha\vert \frac{ \partial \alpha }{\partial k_{x}} \rangle dk_x + \langle  \alpha  \vert \, \frac{ \partial \alpha }{\partial k_{y}}\rangle dk_y\right)
\label{tknn2}
\end{equation}
where the dimensionnal factor $e^2/h$ with $h=2\pi\hbar$ has been reintroduced \cite{thoul} .

The above expression has a well-known topological interpretation: it is the integration of Berry's curvature $\mathcal{F}= \langle \frac{\partial  \alpha }{\partial k_{\mu}} \vert \, \frac{ \partial \alpha }{\partial k_{\nu}} \rangle \, dk^{\mu} \land dk^{\nu}$ over the base space $\mathbb{T}^2$ of a principal fiber bundle $F(\mathbb{T}^2,U(1))$. The number ${\rm Ch}_1 = \frac{i}{2 \pi} \int _{\mathbb{T}^2}
\mathcal{F} $ is a topological invariant referred as \emph{Chern number} and is necessarily an integer. 
Thouless \emph{et al.} showed that  the Hall conductivity \eqref{tknn2} is given by $\sigma _{xy} = \frac{e^2}{h} \sum_{\epsilon _{\alpha} < \epsilon _{\rm F}} \, (t_r - t_{r-1})$ for Hofstadter's model, where $t_r$ is the solution of a diophantine equation
\begin{equation}
r=q \, s_r +p \, t_r
\label{dio}
\end{equation} 
which gives the $r^{\rm th}$ gap's position of the Hofstadter spectrum in terms of the integers $p$ and $q$ of the magnetic flux $\Phi=p/q$. When the Fermi level is exactly situated in the $r^{\rm th}$ gap, one finds that $\sigma_{xy}= t_r \, e^2/h$. The Chern number $t_r$ being an integer one gets the quantization law observed by von Klitzing \cite{klitzing}. 

Now, we show that this result extends to quantum graphs. We suppose that the wave functions accumulate the phase $u_{\alpha} \sim \exp(i\theta_{\alpha})$ along the border $\partial\mathbb{T}^2$ of the torus where the line integral of Eq. (\ref{tknn2}) is carried out, in which case one can write \eqref{tknn2} in the following way
\begin{equation}
\sigma_{x y} =  \frac{e^2}{2 \pi h} \sum_{\rm occ. \, bands}  \oint_{\partial\mathbb{T}^2} \left(\frac{\partial
\theta_{\alpha}}{\partial k_{x}} dk_x+\frac{\partial
\theta_{\alpha}}{\partial k_{y}} dk_y\right)
\label{stokes}\end{equation}

In the weak-coupling limit $r\to 0$, the anisotropy parameter (\ref{Lambda}) is large and the generalized Harper equation (\ref{superharp}) reduces to the approximate eigenvalue equation
\begin{equation}
\cos kl_y \simeq \cos\left( 2\pi\Phi n- k_yl_y\right)
\end{equation}
with $n\in\mathbb{Z}$.  These equations form crossing curves in the plane of the wave number $k$ versus the Bloch parameter $k_y$.  The crossings are exact if $r=0$, but avoided crossings exist if $r$ is small but non vanishing.  In this case, it is possible to calculate the phase $\theta_{\alpha}$ accumulated by the eigenstate $u_{\alpha}$ at each avoided crossing, in a way similar to the one shown for tight-binding models by Kohmoto \cite{komo}.  We obtain that the eigenstates $u_{k_x k_y}$ undergo the transformation 
\begin{equation}
u_{k_x k_y} \rightarrow u_{k_x k_y} \,   \left[- \, e ^{i q k_x  l_x (t_{r}-t_{r-1})} \right]
\end{equation}
when a loop is performed around the first Brillouin zone. One eventually finds  that the Hall conductance is given by
\begin{align}
\sigma _{xy} &=\frac{e^2}{2 \pi h} \sum_{\epsilon _{\alpha} < \epsilon _{\rm F}} \underbrace{\int_{-\frac{\pi}{q l_x}}^{\frac{\pi}{q l_x}} d k_x \frac{\partial \theta}{\partial k_x}}_{2 \pi (t_r - t_{r-1})} \notag \\
&= \frac{e^2}{h} \sum_{\epsilon _{\alpha} < \epsilon _{\rm F}} (t_r - t_{r-1}) \notag \\
&= \frac{e^2}{h} t_r
\end{align}
where we have supposed the Fermi level $\epsilon _{\rm F}$ in the $r^{\rm th}$ gap. \\

\subsection{Phase diagrams of transverse conductivity}

The phase diagram which describes the integer quantum Hall effect for Hofstadter's model has been introduced by Osadchy and Avron \cite{osa}. This diagram shows in a beautiful fractal structure the value of the  Hall conductivity as a function of the magnetic flux and the Fermi energy of the system. Such a phase diagram can be drawn for quantum graphs as well. The result is straightforward for the \emph{self-dual} case $l_x=l_y$ \cite{aubry,mont}, for which the diagram is a nonlinear deformation of the one obtained by Osadchy and Avron \cite{osa}. In order to get the phase diagram of an arbitrary graph, we have solved numerically the generalized Harper equation and resolved the diophantine equation \eqref{dio} for each gap of the spectrum. The phase diagram for the case $l_x=3$ and $l_y=1$ is drawn in Fig. \ref{diagraphin} as a function of the Fermi wave number $k_{\rm F}=\sqrt{2 \epsilon _{\rm F}}$ and the magnetic flux $\Phi$. The quantum phases correspond to the different integer values of the Hall conductivity computed inside the gaps.

\begin{widetext}
\begin{center} 
\begin{figure}%[!b] 
\begin{center} 
{\scalebox{5.1}{\includegraphics{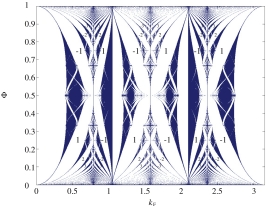}}} 
\caption{\label{diagraphin} Spectrum and phase diagram for a $2D$ quantum graph characterized by $l_x=3$, $l_y=1$, and $r=1$. The horizontal axis is the Fermi wave number $k_{\rm F}=\sqrt{2  \epsilon _{\rm F}}$ and the vertical axis the magnetic flux $\Phi$.  The different phases correspond to the integer values of the Hall conductivity computed inside the gaps. As in the situation depicted in Fig. \ref{graphin}(c) we have $l_x=n$ and $l_y=1$, with $n=3$, and we find exactly $n=3$ butterfly-like structures for $k \in [0, \pi ]$.  
} 
\end{center}
\end{figure} 
\end{center} 
\end{widetext}

A phase diagram describing the IQHE is also obtained for the three-dimensional system described in Sec. \ref{3D}. The Chern numbers evaluated inside the numerous gaps of the initial butterfly ($\gamma=0$) are maintained while the spectrum undergoes the transformation $\gamma \rightarrow 1$. We draw the phase diagram corresponding to the $3D$ quantum graph in Fig. \ref{3Ddia}.

\begin{widetext}
\begin{center} 
\begin{figure}%[!b] 
\begin{center} 
{\scalebox{4.5}{\includegraphics{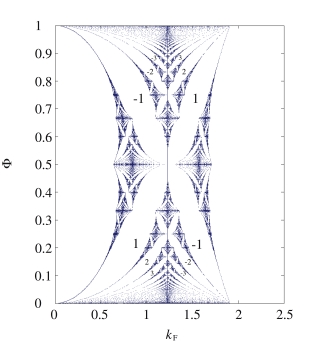}}} 
\caption{\label{3Ddia} Spectrum and phase diagram for a $3D$ quantum graph characterized by $l_x=, l_y=l_z=1$.  The horizontal axis is the Fermi wave number $k_{\rm F}=\sqrt{2 \epsilon _{\rm F}}$ and the vertical axis the magnetic flux $\Phi$.  The different phases correspond to the integer values of the Hall conductivity computed inside the gaps.} 
\end{center}
\end{figure} 
\end{center} 
\end{widetext}

\section{Conclusion}
\label{conclusions}

In this article an important aspect of the quantum transport on $2D$ and $3D$ graphs have been studied namely the quantization of the system's Hall conductivity. 

First, we have obtained the energy spectra of quantum graphs without and with magnetic field.  We showed that their eigenvalue equation can be mapped onto a generalized Harper equation in the case of a $2D$ rectangular lattice.  A $3D$ rectangular lattice has also been considered.  

In zero magnetic field, the graphs have continuous band spectra because of their spatial periodicity and the spectra are discrete at fixed values of Bloch's parameters.  The $2D$ and $3D$ graphs are shown to differ by their level spacing statistics.  Indeed, the $2D$ graph has at most two incommensurate bond lengths so that its level spacing distribution typically presents a gap around zero spacing.  In contrast, the $3D$ graph has at most three incommensurate bond lengths which is sufficient for Wigner repulsion to manifest itself in the level spacing statistics.  On the other hand, both the $2D$ and $3D$ graphs are classically chaotic with a positive Kolmogorov-Sinai entropy per unit time in the classical limit.

In non-zero magnetic field, we have obtained fractal energy spectra.  A deformed version of Hofstadter's butterfly is recovered in the case of a $2D$ lattice with the $C_4$ square symmetry.  If the lattice becomes anisotropic, some gaps are filled and the corresponding zones darken in the spectrum due to the appearance of continuous parts in the spectrum.  Nevertheless, other gaps remain which are characterized by Chern's topological quantum numbers.  We show that the transverse conductivity is quantized in terms of Chern's numbers, as in the integer quantum Hall effect.  We construct the fractal quantum phase diagrams of the transverse conductivity.

In conclusion, quantum graphs show rich structures such as fractal spectra and reveal interesting quantum properties such as those emphasized in this work. These versatile models are promising for the exploration of quantum phenomena in condensed matter systems such as the quantum Hall effects. \\

{\bf Acknowledgments.} N. G. thanks the F.~R.~I.~A. and the F.~R.~S.- F.~N.~R.~S. for financial
support. This research is financially supported by the
''Communaut\'e fran\c caise de Belgique'' (contract ''Actions de Recherche Concert\'ees''
No. 04/09-312) and the F.~R.~S.-FNRS Belgium (contract F. R. F. C. No.
2.4542.02).

\end{document}